\documentclass[sigconf]{acmart}

\newcommand{\fig}[1]{Figure~\ref{fig:#1}}

\newcommand{\tion}[1]{\S\ref{sec:#1}}

\usepackage{balance}
\usepackage{multirow}
\usepackage{blindtext, graphicx}
\usepackage{boldline}
\usepackage{subfig} 
\usepackage{booktabs}
\usepackage{ragged2e}
\usepackage{dblfloatfix} 
\usepackage[para,online,flushleft]{threeparttable}
\usepackage{lipsum}

\usepackage[tikz]{bclogo}
\newenvironment{RQ}[1]%
{\noindent\begin{minipage}[c]{\linewidth}%
\begin{bclogo}[couleur=gray!20,%
                arrondi=0.1,%
                logo=\bctrombone,%
                ombre=true]{~#1}}%
{\end{bclogo}\end{minipage}\vspace{2mm}}

\newcommand{\bi}{\begin{itemize}}
\newcommand{\ei}{\end{itemize}}

\usepackage{enumitem}
\setlist[itemize]{leftmargin=*}
\setlist[enumerate]{leftmargin=*}
\usepackage{makecell}
\usepackage[linesnumbered,ruled,vlined]{algorithm2e}
\setlist{nolistsep} 

\SetKwProg{Fn}{Function}{}{}

\setlist[1]{itemsep=0pt}

\usepackage{amsmath}

\usepackage{times}
\usepackage{colortbl}
\usepackage{tikz}

\makeatletter
\let\th@plain\relax
\makeatother

\definecolor{Gray}{rgb}{0.88,1,1}
\definecolor{Gray}{gray}{0.85}
\definecolor{lightgray}{gray}{0.8}
\usepackage[framed]{ntheorem}
\usepackage{framed}
\usepackage{tikz}
\usetikzlibrary{shadows}
\theoremclass{Lesson}
\theoremstyle{break}

\tikzstyle{thmbox} = [rectangle, rounded corners, draw=black,
fill=Gray!20,  drop shadow={fill=black, opacity=1}]

\newshadedtheorem{lesson}{Answer to RQ}

\renewcommand{\textrightarrow}{$\rightarrow$}

\setcopyright{rightsretained}

\acmConference[NL4SE@ ESEC/FSE 2018]{Workshop on NLP for Software Engineering}{4th November, 2018}{Lake Buena Vista, Florida}
\acmYear{2018}
\copyrightyear{2018}

\acmPrice{15.00}

\setcopyright{usgovmixed}

\acmDOI{0000001.0000001}

\received{August 2018}

\begin{document}
\title[Total Recall and SE]{Total Recall, Language Processing, and  Software Engineering}

\author{Zhe Yu, Tim Menzies}
\affiliation{%
  \institution{North Carolina State University, USA}}
\email{zyu9@ncsu.edu, timm@ieee.org}

\renewcommand{\shortauthors}{Z.Yu, T.Menzies}

\begin{abstract}
A broad class of software engineering problems can be generalized as the ``total recall problem''. This short paper claims that identifying and exploring   total recall language processing problems in software engineering is an important task with wide applicability.

To make that case, we show that by applying and adapting the state of the art active learning and text mining, solutions of the total recall problem, can help solve two important software engineering tasks: (a)~supporting large literature reviews and (b)~identifying software security vulnerabilities. Furthermore, we conjecture that (c)~test case prioritization and (d)~static warning identification can also be categorized as the total recall problem. 

The widespread applicability of ``total recall'' to software engineering suggests that there exists some underlying framework that encompasses not just natural language processing, but a wide range of important software engineering tasks.

\end{abstract}

%
%

%
%

\maketitle


\section{Introduction}
\label{sec:introduction}

Software engineering is a discipline closely involved with human activities. How to help software developers produce better software more efficiently is the core topic of software engineering. Prioritizing tasks can efficiently reduce the human efforts and time required to achieve certain goals of software development. Many of such prioritization problems in software engineering can be generalized as the 
{\em total recall problem} (defined in the next section). 

The total recall problem has been explored in information retrieval for years, and the state of the art solution with active learning and natural language processing aims to resolve the following challenges:
\bi
\item
Among a finite number of tasks, which task to be executed first so that certain goals can be achieved earlier?
\item
At what point, there is no need to execute the remaining tasks?
\item
Are all the tasks executed correctly? How to identify wrongly executed tasks and correct them?
\item
How to scale up the process with multiple humans working in parallel (e.g. crowdsourcing)?
\ei
The first challenge has been extensively explored~\cite{cormack2014evaluation,wallace2010semi,miwa2014reducing} while the rest three have much room to improve~\cite{Cormack2016Engineering,Cormack2017Navigating,cormack2016scalability}. 

This short paper claims that identifying and exploring the total recall problems in software engineering is an important task which benefits both software engineering research and the general solution to total recall problems. To support this claim, in \tion{The Total Recall Problem} we first introduce the general total recall problem, and its state of the art solutions; then in \tion{Total Recall Problems in Software Engineering}, we provide two software engineering tasks which are solved by applying total recall techniques and better solutions to the challenges of total recall problems were created during the process~\cite{Yu2018,DBLP:journals/corr/YuM17,DBLP:journals/corr/abs-1803-06545}. We also conjecture that two other software engineering tasks can be categorized as the total recall problem. The goal of this short paper is to inspire software engineering researchers to explore total recall problems in software engineering with our preliminary results.

\section{The Total Recall Problem}
\label{sec:The Total Recall Problem}

\begin{figure}[!t]
    \centering
    \includegraphics[width=0.45\textwidth]{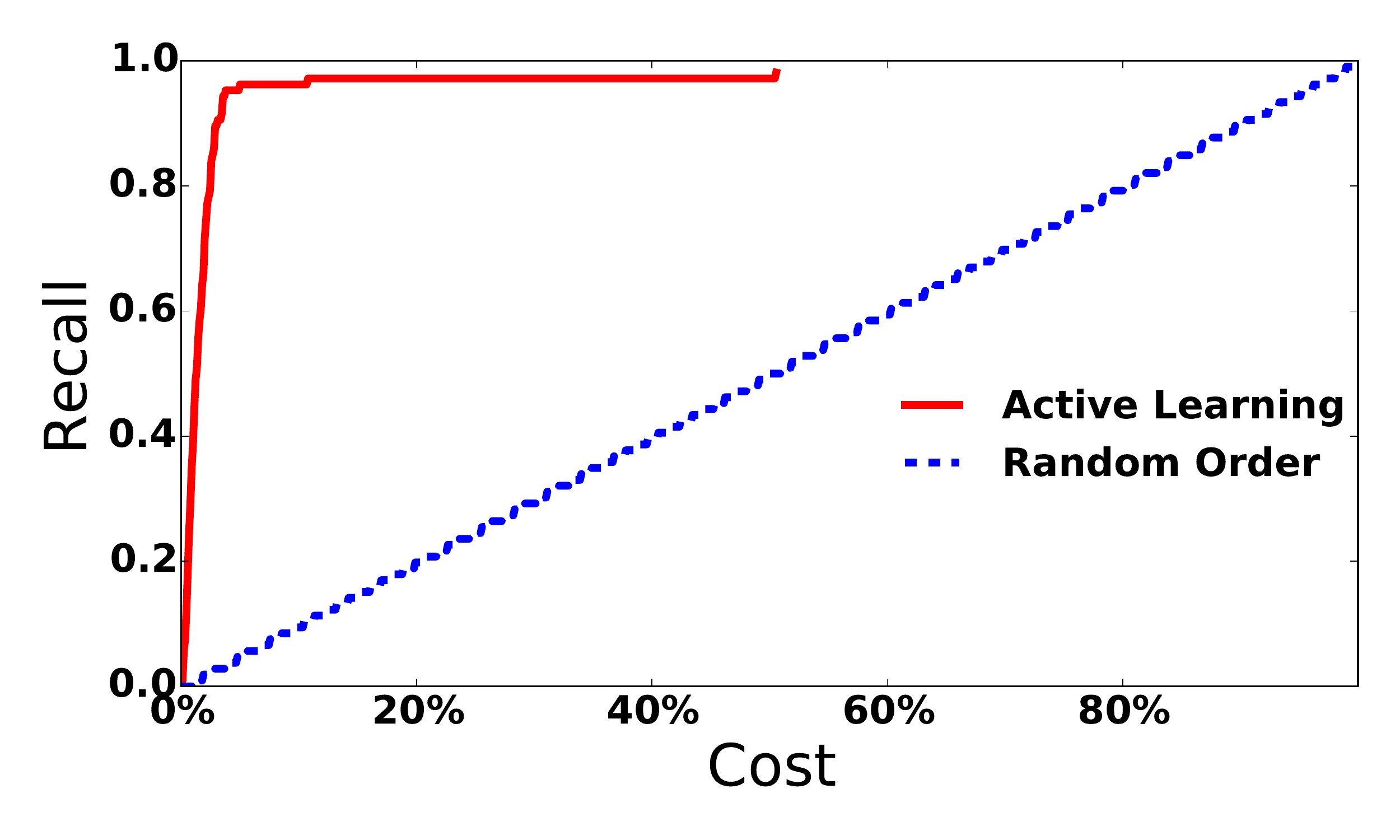}
    \caption{A demonstration for the power of active learning on the total recall problem. X axis shows the cost spent while Y axis shows the recall reached. The dashed line represents the retrieval curve with random inspection order and the solid line shows the retrieval curve with active learning selecting which example to inspect next.}
    \label{fig:demo}
\end{figure}

The total recall problem in information retrieval aims to optimize the cost for achieving very high recall--as close as practicable to 100\%--with a human assessor in the loop~\cite{grossman2016trec}. More specifically, the total recall problem can be described as following:

\begin{RQ}{The Total Recall Problem:}
   Given a set of candidate examples $E$, in which only a small fraction $R \subset E$ are positive, each example $x$ can be inspected to reveal its label as positive ($x\in R$) or negative ($x \not\in R$) at a cost. Starting with the labeled set $L = \emptyset$, the task is to inspect and label as few examples as possible ($\min |L|$) while achieving very high recall $|L\cap R|/|R|$.
\end{RQ}
Different strategies have been applied to solve the total recall problem, including supervised learning and semi-supervised learning. However, the state of the art solutions to the total recall problem apply active learning~\cite{grossman2016trec} to learn from natural language processing features (e.g. bag-of-words, tf-idf) extracted from the current labeled set $L$ and re-rank the rest of the candidate examples $E \setminus L$ so that examples that are more likely to be positive get inspected next. 
This active learning solution has been proven effective in different use cases. A demonstration of the benefit of this active learning strategy can be found in Figure~\ref{fig:demo}, where with active learning (solid red line), high recall (close to 100\%) can be achieved by inspecting only a small portion of the candidate examples.

In evidence-based medicine, researchers screen titles and abstracts to determine whether one paper should be included in a certain systematic review. Wallace et al.~\cite{wallace2010semi} designed patient active learning to help researchers prioritize their effort on papers to be included. With patient active learning, half of the screening effort can be saved while still including most relevant papers~\cite{wallace2010semi}.

In electronic discovery, attorneys are hired to review massive amount of documents looking for relevant ones for a certain legal case and provide those as evidences. Cormack and Grossman~\cite{cormack2014evaluation} proposed continuous active learning to save attorneys' effort from reviewing non-relevant documents, which further can save a large amount of the cost of legal cases. 

The active learning strategy for total recall problem can be described as following:
\begin{enumerate}[start=0,label={\bfseries Step \arabic*}]
\item
Given a candidate set $E$ and initialize the labeled set $L = \emptyset$.
\item
\label{step 2}
Using strategies like ad hoc search or random sampling to select next example $x$ to label ($L\leftarrow L \cup x$).
\item
\label{step 3}
Repeat \ref{step 2} until ``enough'' positive examples have been labeled ($|L \cap R|\ge K$).
\item
\label{step 4}
Train/update a supervised learning model with current labeled set $L$.
\item
\label{step 5}
Use the trained model to predict on the unlabeled set $E\setminus L$ and select next example $x$ to label  ($L\leftarrow L \cup x$).
\item
\label{step 6}
Repeat \ref{step 4} and \ref{step 5} until a stopping rule has been met.
\end{enumerate}
Where detail settings vary from case studies to case studies: 
\bi
\item
In \ref{step 3}, Cormack and Grossman~\cite{cormack2014evaluation} believe that learning should start as soon as possible ($K=1$) while Wallace et al.~\cite{wallace2010semi} suggest to start learning until more training examples are available.
\item
In \ref{step 4}, most studies exact text features from examples and train a support vector machine with linear kernel. However, there are different opinions on how to balance the training data, e.g. Wallace et al.~\cite{wallace2010semi} proposed a technique called aggressive undersampling to drop off the negative examples closes to the positive ones while Miwa et al.~\cite{miwa2014reducing} adjusted the weight of each training example to punish more for misclassifying positive examples.
\item
In \ref{step 5},  Cormack and Grossman~\cite{cormack2014evaluation} select examples that are most likely to be positive while  Wallace et al.~\cite{wallace2010semi} select most uncertain examples.
\item
In \ref{step 6}, Cormack and Grossman~\cite{cormack2014evaluation} stop the process when  a sufficient number of positive examples have been found while Wallace et al.~\cite{wallace2010semi} stop the learning when the model becomes stable and then apply the model to classify the unlabeled examples.
\ei

While this active learning and natural language processing strategy has been extensively explored to resolve the main challenge of total recall problems, there still exists large room for improvement in the three other challenges.

\subsection{When to Stop}
\label{sec:When to Stop}

In practice, there is no way to know the number of positive examples before inspecting and labeling every candidate example. It is thus impossible to know exactly what recall has been reached during the process. Then, how do we know when to stop if the target is reaching, say 95\% recall? This is a practical problem that directly affects the usability of the active learning strategy. Stopping too early will result in missing many valuable positive examples; while stopping too late will cause unnecessary cost when there are no more positive examples to retrieve. So far, researchers have developed various stopping rules to solve the ``when to stop'' problem. 
\bi
\item
Ros et al.~\cite{ros2017machine} developed the the most straightforward rule, which decides that the process should be stopped after 50 negative examples are found in succession. 
\item
Cormack and Grossman proposed the knee method~\cite{Cormack2016Engineering}, which detects the inflection point $i$ of current recall curve, and compare the slopes before and after $i$. If $slope_{<i}/slope_{>i}$ is greater than a specific threshold $\rho$, the review should be stopped.
\item
Wallace et al.~\cite{wallace2013active} applied an estimator to estimate the number of positive examples $|R|$ and let the users decide when to stop by showing them how close they are to the estimated number.
\ei

\subsection{Human Error Correction}
\label{sec:Human Error Correction}

When solving the total recall problems, the next example to be inspected relies on model trained on previously labeled examples. What if those labels can be wrong? The trained model could be misled by wrongly labeled examples and thus make worse decision on which example to inspect next.
By now, researchers have applied various strategies to correct such human errors:
\begin{itemize}
\item
One simple way to correct human errors is by majority vote~\cite{Kuhrmann2017On}, where every example will be inspected by two different humans, and when the two humans disagree with each other, a third human is asked to make the final decision.
\item
Cormack and Grossman~\citep{Cormack2017Navigating} built an error correction method upon the knee stopping rule~\citep{Cormack2016Engineering}. After the review stops, examples labeled as positive but are inspected after the inflection point ($x>i$) and examples labeled as negative but inspected before the inflection point ($x<i$) are sent to humans for a recheck.
\end{itemize}


\subsection{Scalability}

How to scale up the active learning framework with multiple humans working simultaneously on the same project remains an open challenge to the total recall problem. 
In electronic discovery, Cormack and Grossman~\cite{cormack2016scalability} proposed S-CAL where model trained on a finite number of examples is used to predict and estimate the precision, recall, and prevalence of the target large set of examples. However, this approach sacrifices the adaptability of the active learner. In evidence-based medicine, Nguyen et al.~\cite{nguyen2015combining} explored the task allocation problem where two types of human operators are available: (1) crowdsourcing workers who are cheaper but less accurate, and (2) experts who are much more expensive but also more accurate. This approach allows the system to operate more economically but still cannot scale up with growing number of training examples.  

\section{Total Recall Problems in Software Engineering}\label{sec:Total Recall Problems in Software Engineering}

As discussed in \tion{introduction}, there are tasks in software engineering that resemble the problem of total recall. In this section, we will discuss four such problems. The first two tasks have been explored by the authors previously and the promising results convince us that applying active learning and natural language processing strategy is the key to solving these software engineering problems.

\subsection{Primary Study Selection in Systematic Literature Reviews}

For the same reason of medicine researchers, software engineering researchers also need to read literature to stay current on their research topics. Researchers conduct systematic literature reviews~\cite{kitchenham2004evidence} to analyze the existing literature and to facilitate other researchers. Among different steps of systematic literature reviews, primary study selection is in exactly the same format as citation screening in medical systematic reviews. The problem is also for reading titles and abstracts to find relevant papers to include, except that those papers are about software engineering. As a result, the primary study selection problem can be described as a total recall problem:
\bi
\item 
$E$: set of software engineering papers returned from a search query.
\item
$R$: set of relevant papers to the systematic literature review.
\item
$L$: set of papers that have been reviewed and labeled as relevant or non-relevant by human researchers.
\ei

While exploring this total recall problem~\cite{Yu2018,DBLP:journals/corr/YuM17}, the authors:
\begin{enumerate}
\item
Designed a new active learning framework by combining advantages of the state of the art approaches~\cite{wallace2010semi,miwa2014reducing,cormack2014evaluation} in \tion{The Total Recall Problem}. When reaching the same recall, the new framework costs 20-50\% less than these prior state of the art approaches.
\item
Created an estimator for $|R|$ based on semi-supervised learning for the \emph{when to stop} challenge. This estimator showed better accuracy than the one from Wallace et al.~\cite{wallace2013active} and provided better stopping rules than the prior methods in \tion{When to Stop}. With the help of this estimator, the users can now know what level of estimated recall has been reached and make better decision on whether to spend more effort to achieve a higher recall at any point of the process.
\item
Proposed an error identification method, in which only examples with labels that the current active learner disagrees most on are rechecked by a different human expert. This method has been proven to be more cost effective in addressing the \emph{human error correction} challenge than prior methods in \tion{Human Error Correction}.
\end{enumerate}

Datasets and tools for reproduction and making further improvements on the primary study selection problem can be found at Seacraft, Zenodo\footnote{https://doi.org/10.5281/zenodo.1162952} and Github\footnote{https://github.com/fastread/src}.

\subsection{Software Security Vulnerability Prediction}
\label{sec:Software Security Vulnerability Prediction}

\begin{figure}[!t]
    \centering
    \includegraphics[width=0.9\linewidth]{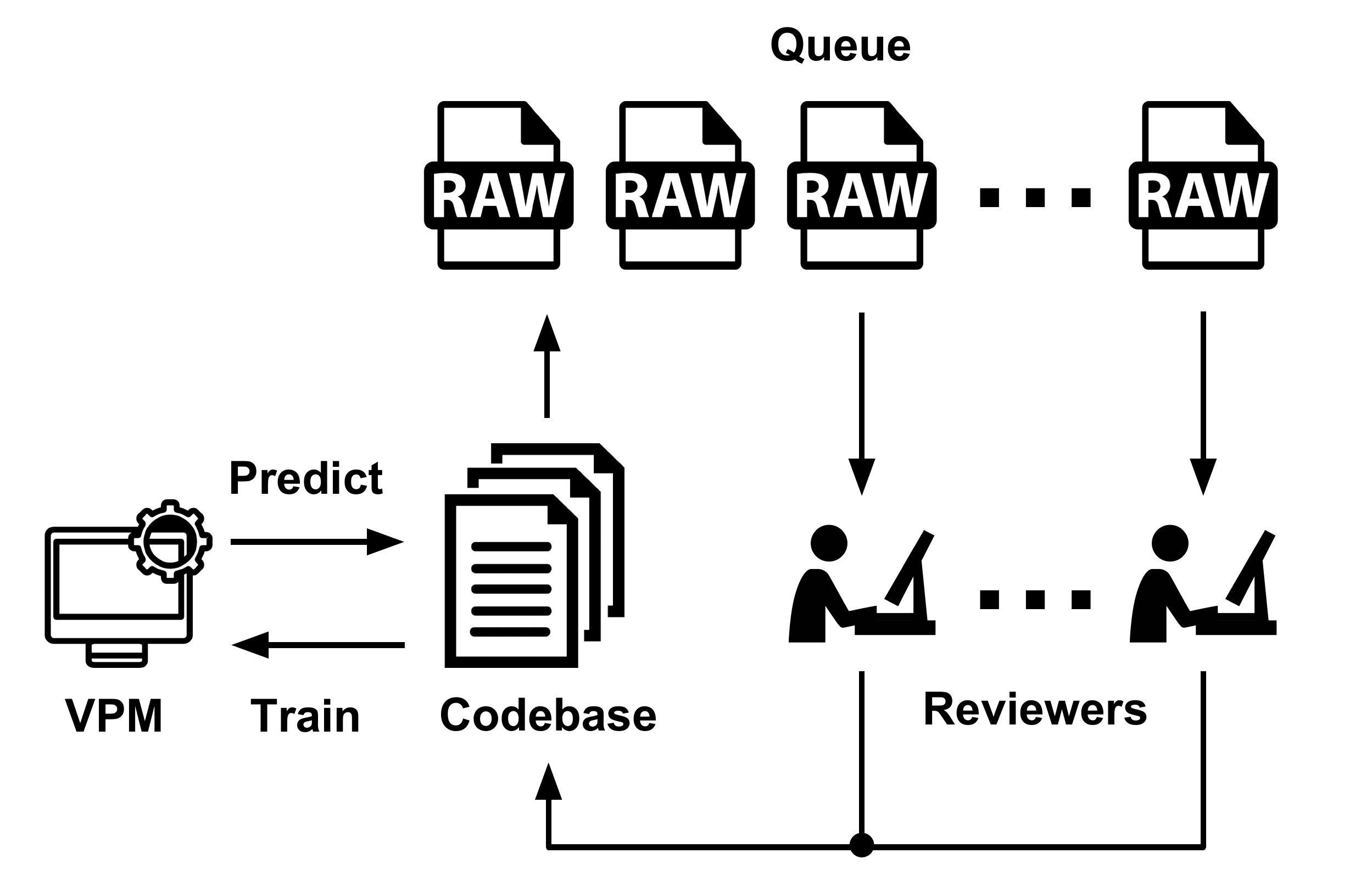}
    \caption{Poactive Security Review and Test Framework}
    \label{fig:distribute}
\end{figure}

Society needs more secure software. It is crucial to identify software security vulnerabilities and fix them as early as possible. However, it is time-consuming to have human experts inspecting the entire codebase looking for the few source code files that contain vulnerabilities. The solution to this problem, prioritizing the human inspection effort towards codes that are more likely to have vulnerabilities is also a total recall problem:
\bi
\item 
$E$: the entire codebase of a software project.
\item
$R$: set of source code files that contain vulnerabilities.
\item
$L$: set of source code files     already   inspected by humans.
\ei

With simulations on the known vulnerabilities from Mozilla Firefox project~\cite{DBLP:journals/corr/abs-1803-06545}, the authors:
\begin{enumerate}
\item
Extracted bag-of-words from source codes and applied same active learning algorithm as the primary study selection problem~\cite{Yu2018,DBLP:journals/corr/YuM17} to select which source code file to inspect next and to stop the inspection when target recall is reached. Results showed that 95, 99, 100\% recall can be achieved by having human experts inspect 28, 41, 55\% of the source code files, respectively~\cite{DBLP:journals/corr/abs-1803-06545}. Results also showed that text features perform better than traditional software metrics features.
\item
Adapted the error correction method from primary study selection~\cite{DBLP:journals/corr/YuM17} to identify missing vulnerabilities\footnote{When inspecting codes, human may miss some vulnerabilities (false negatives) but may never have false positives, and the false negative rate can be as high as 47\%~\cite{hatton2008testing}.}. Results showed that it can reach higher recall with lower cost comparing to other error correction methods mentioned in \tion{Human Error Correction}.
\item
Designed a centralized system where human operators can work in parallel but all the information is gathered in one place to update the model, as shown in \fig{distribute}. However, when the number of training example grows, the time required for updating the model becomes longer and longer. 
\end{enumerate}
The real benefit of treating vulnerability prediction as a total recall problem is that no labeled data (known vulnerabilities) are required to start the process, thus these vulnerabilities can be identified and fixed before the software is deployed. On contrast, conventional methods such as supervised learning or unsupervised learning require post-deployment information such as known vulnerabilities from bug reports or crash dump stack traces.

\subsection{Static Warning Identification}

Static Analysis tools (e.g., FindBugs) are widely used
to find defects in software. These tools scan source code or binary
code of a software project and infer bugs, security vulnerabilities, and bad programming practices with heuristic pattern matching techniques~\cite{wang2018}. Problem is, the high false positive rate of the reported warnings causes most of the warnings not acted
on by developers~\cite{wang2018}. Reducing such false positives can be viewed as a total recall problem:

\bi
\item 
$E$: all warnings reported by the static analysis tools.
\item
$R$: set of warnings that reveal true defects.
\item
$L$: set of warnings that have been inspected by human experts.
\ei

This static warning identification problem has been explored with supervised learning methods and sets of different features~\cite{wang2018}. Analyzing the warnings with natural language processing and applying active learning to select which warning to inspect might help reduce false positives and make static analysis tools more practical to use.




\subsection{Test Case Prioritization}
\label{sec:Test Case Prioritization}

Regression testing is an expensive testing process to detect whether new faults have been introduced into previously tested code. To reduce the cost of regression testing, software testers may prioritize their test cases so that those which are more likely to fail are run earlier in the regression testing process~\cite{elbaum2002test}. The goal of test case prioritization is to increase the rate of fault detection, and by doing so, it can provide earlier feedback on the system under test, enable earlier debugging, and increase the likelihood that, if testing is prematurely halted, those test cases that offer the greatest fault detection ability in the available testing time will have been executed~\cite{elbaum2002test}. This test case prioritization problem can also be treated as a total recall problem with the following notations:
\bi
\item 
$E$: all candidate regression test suites.
\item
$R$: set of test suites that will detect faults.
\item
$L$: set of test suites that have been executed.
\ei

Existing techniques for prioritizing test cases are ``unsupervised'', i.e. these techniques decide an order of the test cases to be run and stick with it. Applying active learning to adapt the ordering with knowledge learned from test cases executed can potentially further increase the rate of fault detection and reduce more cost. However, this problem has not been explored as a total recall problem and the following challenges need to be resolved:
\bi
\item 
What types of feature can be extracted from the test cases so that the learned model can accurately predict the likelihood of fault detection from a test case before its execution. 
\item
How to balance the priority of test cases increasing coverage of statements/functions most and the more-likely-to-fail test cases.
\item
How to handle test dependence, i.e.  if test cases are not independent, changing their order may fail some tests but pass others~\cite{zhang2014empirically}.
\ei

\section{Conclusions and Future work}

Many of the software engineering problems can be generalized as the total recall problem. This paper identified four such problems. With two case studies of primary study selection and vulnerability prediction, we showed that exploring these total recall problems in software engineering would benefit both software engineering research and the general solution to total recall problems. We hope this paper will attract more researchers studying and improving the total recall problems in software engineering. Future work in this area includes but is not limited to the following:
\bi
\item
Improve core algorithm for the total recall problem---reaching same recall with less cost.
\item
Build more accurate estimator for current recall achieved.
\item
Resolve human errors more efficiently.
\item
Scale up the total recall solutions (probably by ensemble learning) and utilize low-cost workers through crowdsourcing.
\item
Apply total recall techniques to reduce false alarms in static code analysis and to prioritize test cases.
\item
Identify more total recall problems in software engineering.
\ei

\balance
\bibliographystyle{ACM-Reference-Format}

\end{document}